# Fast 3D Ultrasound Localization Microscopy via Projection-based Processing Framework

Jingke Zhang, Jingyi Yin, U-Wai Lok, Lijie Huang, Ryan M. DeRuiter, Tao Wu, Kaipeng Ji, Yanzhe Zhao, James D. Krier, Xiang-yang Zhu, Lilach O. Lerman, Chengwu Huang*, and Shigao Chen*

*Abstract*—Three-dimensional ultrasound localization microscopy (ULM) enables comprehensive visualization of the vasculature, thereby improving diagnostic reliability. Nevertheless, its clinical translation remains challenging, as the exponential growth in voxel count for full 3D reconstruction imposes heavy computational demands and extensive post-processing time. In this row-column array (RCA)-based 3D *in vivo* pig kidney ULM study, we reformulate each step of the full 3D ULM pipeline, including beamforming, clutter filtering, motion estimation, microbubble separation and localization into a series of computational-efficient 2D operations, substantially reducing the number of voxels to be processed while maintaining comparable accuracy. The proposed framework reconstructs each 0.75-s ensemble acquired at frame rate of 400 Hz, covering a 25×27.4×27.4 mm³ volume, in 0.52 s (70% of the acquisition time) on a single RTX A6000 Ada GPU, while maintaining ULM image quality comparable to conventional 3D processing. Quantitatively, it achieves a structural similarity index (SSIM) of 0.93 between density maps and a voxel-wise velocity agreement with slope of 0.93 and $R^2$ = 0.88, closely matching conventional 3D results, and for the first time, demonstrating potential for real-time feedback during scanning, which could improve robustness, reduce operator dependence and accelerate clinical workflows.

## I. INTRODUCTION

Ultrasound localization microscopy (ULM) is an emerging technique that enables super-resolution imaging of the microvasculature by localizing and tracking intravenously injected microbubbles (MB) as they flow through blood vessels [1, 2]. By accumulating MB trajectories from tens of thousands of frames acquired over several minutes, ULM can achieve up to a tenfold improvement in spatial resolution [3-7]. With its exceptional spatial resolution, 2D ULM has facilitated early diagnosis [8, 9], investigation of physiological changes across disease progression [10-14], and evaluation of therapeutic response [15] both in animal models and in humans.

Advancing from 2D to 3D ULM enables more complete visualization of the vasculature, reduces dependence on user-selected imaging planes, thereby improving its diagnostic reliability [16]. Clinical translation of 3D ULM is expected to greatly benefit disease evaluation in numerous applications [17] [18]. However, implementing 3D ULM in clinical practice remains challenging. Unlike conventional 2D imaging, which reconstructs only a single plane, 3D imaging requires reconstruction of an entire volume, leading to an exponential increase in the number of voxels to be processed. Moreover, because thousands of frames are required to capture sufficient MB trajectories, tens of gigabytes of data are generated, creating a substantial burden on storage, transfer, and processing in clinical workflows[17][19]: reconstruction takes several hours on CPU [20, 21], and still requires tens of minutes even with GPU acceleration [22, 23], primarily because the number of voxels increases cubically with resolution. As a result, every processing step, from clutter filtering and motion estimation to MB localization, must handle substantially more data than in 2D. Take normalized cross-correlation (NCC)-based MB localization as an example: correlating with a 3D Gaussian kernel of size $k$ requires accessing and multiplying $k^3$ voxels per neighborhood, yielding a cost of $\mathcal{O}(k^3 s^3)$ for a volume of size $s$. In contrast, the 2D case scales as $\mathcal{O}(k^2 s^2)$, since each pixel involves a smaller 2D neighborhood, greatly reducing the computational burden by $ks$-fold. Therefore, reformulating 3D operations into a series of 2D operations with minimal loss of accuracy is promising to substantially reduce complexity and enable on-the-fly ULM reconstruction. It would allow immediate visualization of microvasculature and hemodynamics, enabling timely diagnostic and improving clinical robustness.

Row-column array (RCA) has recently emerged as a research hotspot, enabling 3D imaging with substantially fewer channels [24, 25], and have attracted growing interest in ULM [26-31].

Here we introduce a fast projection-based processing framework for 3D ULM that reformulates volumetric operations into a series of 2D steps, reducing reconstruction time from the typical several minutes to 0.52 s for a 0.75 s

The study was partially supported by the National Institute of Diabetes and Digestive and Kidney Diseases under award numbers of R01DK129205 and R01DK138998. The content is solely the responsibility of the authors and does not necessarily represent the official views of the National Institutes of Health. The Mayo Clinic and some of the authors (J. Z., U-W. L., C. H., and S. C.) have pending patent applications related to the technologies referenced in this publication. (Corresponding authors: Shigao Chen and Chengwu Huang)

J. Zhang, J. Yin, UW. Lok, L. Huang, R. M. DeRuiter, K. Ji, Y. Zhao, C. Huang, and S. Chen are with the Department of Radiology, Mayo Clinic College of Medicine and Science, Rochester, MN 55905 USA (e-mail: Chen.Shigao@mayo.edu, Huang.Chengwu@mayo.edu).

T. Wu is with the Department of Medical Ultrasonics, The Third Affiliated Hospital of Sun Yat-Sen University, Guangzhou 510080 China, and the Department of Radiology, Mayo Clinic College of Medicine and Science, Rochester, MN 55905 USA.

X.-Y. Zhu, J. D. Krier, and L. O. Lerman are with the Division of Nephrology and Hypertension, Mayo Clinic, Rochester, MN 55905 USA.

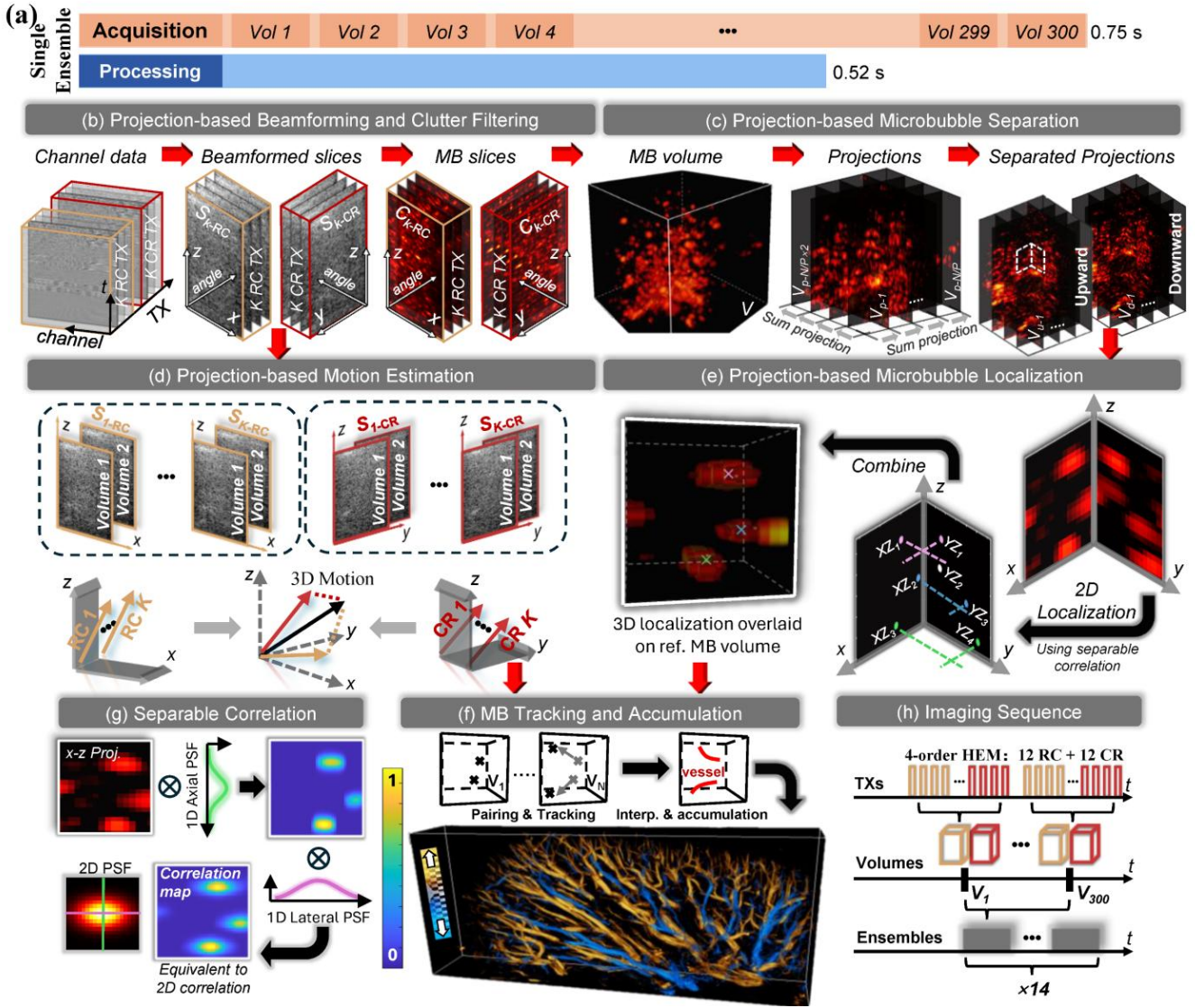

Fig. 1. Overview of the proposed projection-based processing framework for 3D ULM. (a) Acquisition time versus ULM processing time for a 300-volume ensemble of a pig kidney data. (b) Central slices along the transmit directions of the volume are beamformed after each transmission, and clutter filtered to extract MB signals. (c) For each steered RC/CR transmission, a full volume is extrapolated from the central slice. Volumes from all transmissions are compounded, adjacent slices along orthogonal directions are grouped and sum-projected, followed by MB separation. (d) 2D motion estimation was performed on beamformed slices at each angle: $x$–$z$ plane for RC and $y$–$z$ plane for CR, then combined to recover 3D motion. (e) MBs were localized on the two orthogonal separated projections using 2D separable NCC and paired across planes to recover their true 3D positions. (f) The localized MBs were motion-corrected, paired, tracked, interpolated, and accumulated to generate the final ULM volume. (g) Conventional 2D NCC is accelerated using two 1D separable correlations. (h) The imaging sequence consisted of 12 RC and 12 CR transmissions, encoded with 4th-order Hadamard to generate a single volume. Fourteen ensembles of 300 volumes each were acquired.

ensemble on a single RTX A6000 Ada GPU (70% of the acquisition), while maintaining the similar quality to that would have been produced through conventional full 3D processing. This, for the first time, enables faster-than-acquisition 3D *in vivo* ULM reconstruction. The proposed framework was comprehensively validated through simulation, phantom, and *in vivo* pig kidney experiments. Each ULM processing step demonstrated comparable performance between conventional full 3D ULM and proposed fast projection-based 3D ULM methods, and quantitative evaluation in a pig kidney confirmed the high similarity of the reconstructed ULM volumes.

## II. METHODS

### A. Projection-based Processing Framework for 3D ULM

ULM processing chain typically involves beamforming, motion estimation, clutter filtering, MB separation, localization, and tracking. In this section, we describe how each of these steps can be implemented using a series of separable 2D processing to improve computational efficiency, in contrast to conventional full 3D processing. RCA-based 3D ULM was used as an illustrative example. While the beamforming step is specifically tailored for RCA, the framework is adaptable to other probes, for example, projected beamforming [32] can be integrated to enable matrix-array-based real-time 3D ULM. In this work, all the computation was performed under the framework of Compute Unified Device Architecture (CUDA). Figure 1a illustrates the acquisition time and corresponding reconstruction time per ensemble using the proposed fast projection-based 3D ULM.

*1) Beamforming*

*From 3D to 2D processing:* This step is illustrated in Fig. 1b. Volumetric beamforming is computationally demanding because each 3D volume typically contains hundreds of 2D slices to be beamformed. In RCA-based plane wave imaging, there is no transmit focusing. While this lack of focusing results in strong sidelobe artifacts [19], it also provides an useful property: all slices along the transmit direction are approximately identical, and intrinsically the projection along the transmit direction, differing only by a rigid axial shift determined by the transmit steering angle. Leveraging this property, a two-stage beamforming strategy has been proposed [33, 34]. In Stage 1, only a single reference slice $S_{k\text{-RC/CR}}$ ($k = 1, 2, …, K$) of size $N_{x/y} \times N_z$ in the imaging region is beamformed from the raw channel data for each one of $K$ RC and $K$ CR transmissions, and then clutter filtered to $C_{k\text{-RC/CR}}$ as detailed in Section IIA2. In Stage 2, the full volume $V_{k\text{-RC/CR}}$ of size $N_x \times N_y \times N_z$ is reconstructed from $C_{k\text{-RC/CR}}$ by extrapolating along the transmit direction for the $k^{\text{th}}$ steered RC and CR transmission. Finally, the compounded volume $V$ is obtained by coherently summing all $V_{k\text{-RC}}$ and $V_{k\text{-CR}}$ ($k = 1, 2, …, K$). For simplicity, in the following part, we assume that all three dimensions have the same voxel count, denoted by $N$.

*Computational complexity reduction:* According to [33], the computational complexity of beamforming the compounded volume $V$ is reduced from $\mathcal{O}(2N_c N^3 K)$ to $\mathcal{O}(2N^3 K)$, where $N_c$ is the number of channels and $2K$ is the total number of RC and CR transmissions.

*CUDA implementation:* We implemented this approach in CUDA using customized kernels, assigning one thread per voxel [34] in each stage to enable fast IQ beamforming.

*2) Clutter filtering*

*From 3D to 2D processing:* This step is illustrated in Fig. 1b. Singular value decomposition (SVD) is the most widely used technique for MB signals extraction. It stacks each volume as a column of the spatiotemporal matrix $I \in \mathbb{C}^{n \times n_t}$, where $n$ and $n_t$ denote the number of spatial voxels and volumes, respectively, leading to a computational complexity of $\mathcal{O}(nn_t^2)$. Recently, eigenvalue decomposition-based clutter filtering [35, 36] was proposed to accelerate this process by transforming the non-parallelizable SVD into three steps: parallelly computing the covariance matrix $C \in \mathbb{C}^{n_t \times n_t}$ of $I$, performing a lightweight eigenvalue decomposition of $C$, and parallelly projecting to recover blood flow signals. Although the theoretical complexity remains $\mathcal{O}(nn_t^2)$ [36], parallelization offers substantial speedup. Since the computational complexity still scales linearly with $n$, reducing the number of voxels is critical. In this study, we therefore applied eigenvalue-based clutter filtering immediately on 2D slices $S_{k\text{-RC/CR}}$ obtained from Stage 1 beamforming to obtain clutter filtered reference slices $C_{k\text{-RC/CR}}$, followed by Stage 2 beamforming to obtain the final compounded MB volumes $V$.

*Computational complexity reduction:* The voxel count in each 3D compounded volume is $N^3$, whereas the Stage 1 beamformed slices contains $2N^2K$ pixels. This corresponds to a voxel count reduction by a factor of $N/(2K)$, leading to an equivalent reduction in computational cost.

*CUDA implementation:* We implemented this approach in CUDA using cuBLAS and cuSOLVER APIs. Covariance matrices of $2K$ beamformed slices and the final projection step to recover MB signals were computed in batch with "*cublasCgemmBatched*", eigenvalue decomposition was performed with "*cusolverDnSsyevjBatched*".

*3) Motion Estimation*

*From 3D to 2D processing:* This step is illustrated in Fig. 1d. Many studies have shown that when breath-induced motion is minimized, rigid motion correction can achieve a good balance between computational cost and correction performance [7, 38]. In this study, sub-pixel motion was estimated using a Fourier-domain normalized cross-correlation (NCC) approach [39]. In brief, each frame is transformed into the Fourier domain, the correlation function between the two frames is computed, and the displacement versus the reference frame is determined by locating the subpixel peak of the correlation function. This procedure requires two forward and one inverse Fourier transform, which dominate the computational cost.

To reduce computational cost, motion estimation was applied immediately after Stage 1 beamforming, the $K$ RC slices and $K$ CR slices were used separately to estimate motion within the *x-z* plane and the *y-z* plane. $K$ displacement vectors from $K$ RC and $K$ CR transmissions were first averaged, and then the RC- and CR-averaged motions were vector-summed to recover the full 3D motion.

*Computational complexity reduction:*

For volumetric data with $N^3$ voxels, a 3D Fourier transform (or its inverse) has computational complexity of $\mathcal{O}(N^3 \log(N))$. In contrast, applying 2D Fourier transforms to the $2K$ $S_{k\text{-RC/CR}}$ slices reduces the complexity to $\mathcal{O}(2KN^2 \log(N))$, decreasing the computation burden by a factor of $N/(2K)$. Furthermore, the correlation operations scale directly with the number of voxels, so their cost is also reduced under the 2D formulation.

*CUDA implementation:*

Fourier transforms and their inverse were implemented using cuFFT API ("*cufftPlanMany*"). The Fourier domain correlation was performed with a customized kernel.

*4) MB Localization*

*From 3D to 2D processing:* This step is illustrated in Fig. 1e. Instead of localizing MBs directly in 3D space using 3D processing, we first project the MB volume along the lateral (*x*) and elevational (*y*) directions to compress the 3D MBs into two projected 2D images (*y-z* and *x-z*). MB localization is then performed independently on each 2D projection to obtain the (*y, z*) and (*x, z*) coordinates, from which the 3D MB positions are reconstructed. To prevent excessive MBs from being projected onto a single 2D plane that may cause overlap and compromise localization accuracy, the whole volume can be divided into multiple smaller sub-blocks before projection and localization. Specifically, the original $N \times N \times N$ volume is divided into $N/P \times N/P$ non-overlapping sub-blocks, each of size $P \times P \times N$. Each sub-block is then projected onto two orthogonal planes (*x-z, y-z*), which together retain the essential spatial information of the original 3D block. Beamforming, clutter filtering and separation steps were applied to the RC and CR volumes separately, and the filtered RC and CR stacks were cross-correlated with a window size of 3 using the XDoppler scheme [40] to enhance the MB signal prior to localization step.

In our conventional 3D localization, the MB volume is first normalized cross-correlated with the system's Gaussian-fitted 3D PSF, and sub-pixel regional peaks in the resulting 3D correlating map are identified as 3D localizations. In our projection-based 3D localization, the two projections are first normalized cross-correlated with the corresponding Gaussian-fitted 2D PSFs (*x-z* and *y-z*), yielding two sets of 2D localizations, $XZ_i$ ($i = 1, 2, …, I$) and $YZ_j$ ($j = 1, 2, …, J$), where $I$ and $J$ indicate the total number of localizations within the XZ and YZ projections, respectively. 3D MB localizations were reconstructed by pairing 2D coordinates obtained from different projected planes. For each $XZ_i$ localization, candidate pairs with all $YZ_j$ localizations were searched within an empirically defined axial tolerance of one voxel. Among all possible pairs, only those whose pixel intensity ratios between 2D localizations fell within the range of 0.9–1.11 were retained. Final pairings were determined by selecting those with the mutual minimum intensity ratio. For each matched pair ($XZ_i$, $YZ_j$), the 3D position was reconstructed by combining the $x$ coordinate from $XZ_i$, the $y$ coordinate from $YZ_j$, and the average $z$ coordinate of the two. To further accelerate speed of normalized cross-correlation between projected planes and PSF kernel, we introduced separable subpixel correlation, where the 2D operation was further decomposed into successive 1D correlations along each dimension of the *x-z* and *y-z* projection planes, as shown in Fig. 1g.

*Computational complexity reduction:* The computational complexity of a direct 3D correlation with a correlation kernel of size $d^3$ is on the order of $\mathcal{O}(N^3 d^3)$. In comparison, 2D correlations on the total of $2N/P$ orthogonal projections with a kernel size of $d^2$ has a complexity of $\mathcal{O}(2N^3 d^2/P)$. When applying separable correlation, it is further reduced to $\mathcal{O}(4N^3 d/P)$. 2D and 2D separable correlations reduce the complexity by factors of $dP/2$ and $d^2P/4$, respectively.

*CUDA implementation:* Before localization, the maximum intensity of each volume was obtained via CUDA Thrust reduction for thresholding. Subsequent correlation and 2D localization pairing were performed with customized kernels.

5) Microbubble Separation

*From 3D to 2D processing:* This step is illustrated in Fig. 1c. MB separation improves localization and tracking by separating spatially overlapping MBs into sub-populations according to their spatiotemporal flow dynamics (speed and direction) [37]. This is particularly important in 3D imaging, where the larger point spread function (PSF) increases overlap probability compared to 2D. The method applies a temporal Fourier transform at each voxel, followed by multiple bandpass filters to isolate MBs within different speed ranges, and subsequent inverse Fourier transform recovers the separated MBs subpopulations.

For conventional 3D processing, separation is applied voxel-wise across the entire compounded volumes $V$. In our proposed projection-based processing, the compounded volumes $V$ are first divided into smaller blocks for localization as described earlier. Subsequently, the blocks are sum-projected along lateral ($x$) and elevational ($y$) directions, resulting in projected planes $V_{p\text{-}m}$ ($m = 1, 2, …, N/P \times 2$) forming $N/P$ planes along $y$-$z$ and $x$-$z$ planes separately. Finally, separation is applied on the projected planes pixel-wise to obtain subsets $V_{u\text{-}m}$ and $V_{d\text{-}m}$, corresponding to upward and downward MBs, respectively.

*Computational complexity reduction:* Because separation is independently applied to each voxel's slow-time signal, the computational cost scales linearly with voxel count. The voxel number is $N^3$ in 3D, versus $N^3/P \times 2$ in 2D, reducing complexity by a factor of $P/2$.

*CUDA implementation:* Fourier transform and the inverse were implemented using cuFFT API ("*cufftPlanMany*"). The bandpass filtering was performed with a customized kernel assigning one thread to each frequency component of the voxels.

6) MB Tracking and Accumulation

This step is illustrated in Fig. 1f. The movement of localized MBs was tracked using a bipartite graph-based pairing and tracking algorithm [41]. Only MBs that were consistently tracked across ten consecutive volumes were retained. The resulting tracks were smoothed with least-squares quadratic fitting, and the trajectories exhibiting a change in movement direction greater than 45°, or a change in speed exceeding half of their own mean flow speed were discarded. The retained tracks were then interpolated and accumulated to reconstruct the final microvascular image.

Unlike conventional approaches, which track each MB continuously from its first appearance until disappearance, our method limits tracking to a fixed length of ten frames to enable parallelization. Each localized MB in every volume was assigned to an individual thread, which independently searched for a trajectory of length ten. If successful, the trajectory was preserved, smoothed, and interpolated, enabling efficient and parallelizable tracking. The entire tracking and accumulation pipeline was implemented in CUDA with customized kernels.

B. Animal preparation

The animal study was approved by the Mayo Clinic Institutional Animal Care and Use Committee (IACUC) under protocol A00006785-22. A 10-week-old domestic pig was anesthetized with intramuscular Telazol (5 mg kg$^{-1}$) and Xylazine (2 mg kg$^{-1}$), and maintained with intravenous ketamine (11 mg kg$^{-1}$ h$^{-1}$) and Xylazine (1.8 mg kg$^{-1}$ h$^{-1}$). The animal was intubated and mechanically ventilated with room air, and catheterized via the carotid artery and external jugular vein. To prevent acute allergic reactions to ultrasound contrast agents, diphenhydramine (1 mg kg$^{-1}$) and dexamethasone (2 mg) were given intravenously 30–60 min before MB injection [18]. The pig was positioned in lateral recumbency, and hair over the ultrasound site was shaved.

### C. Data Acquisition and processing for 3D ULM

In pig kidney study, the ultrasound probe was securely clamped during data acquisition to minimize motion. Following a 0.75 ml bolus injection of Definity MB suspension (Lantheus Inc., North Billerica, MA), real-time maximum intensity projection (MIP) along both lateral and elevational directions of the 3D clutter-filtered MB volume were used to guide probe positioning. The kidney was scanned *in vivo* using an RCA transducer (Daxsonics Inc., Halifax, NS, Canada) with pitch size of 0.214 mm, bandwidth of 65%, and 128 + 128 elements (Row + Column), connected to a Verasonics Vantage 256 system (Verasonics, Kirkland, WA, USA). The transmit frequency was set to 5.21 MHz instead of the probe's 7 MHz center frequency, to maximize the signal-to-noise ratio [29].

The imaging sequence is illustrated in Fig. 1h. Four-order Hadamard-encoded orthogonal plane wave (OPW) transmissions [42] with 12 RC and 12 CR steered transmissions using angular pitch of 0.6° were used to acquire 4,200 volumes at a volume rate of 400 Hz. Specifically, 14 buffers of 300 volumes each (0.75 s per buffer) were collected and continuously saved to disk. To reduce data volume, acquisition was restricted to a depth range of 20–45 mm. The mechanical index (MI) was measured using a calibrated hydrophone (Ondacorp, Sunnyvale, CA, USA) in a tank of degassed water as 0.16 at depth of 20 mm with a transmit voltage of 25 V.

In-phase/Quadrature (IQ) beamforming was first performed to reconstruct slices corresponding to each angled transmission with a $\lambda/3 \times 2\lambda/3$ (axial × lateral/elevational; $\lambda$ = 300 μm) pixel size in the first stage. After clutter filtering, they were extrapolated into volumes with a $\lambda/2 \times 2\lambda/3 \times 2\lambda/3$ (axial × lateral × elevational) voxel size (see Section 2A1). ULM density and velocity volumes were subsequently reconstructed on a finer grid with $50 \times 50 \times 50$ μm$^3$ voxel size, covering a region of $25 \times 27.4 \times 27.4$ mm$^3$. In the proposed projection-based 3D ULM processing, the size $P$ of sub-block for projection was set to 8. In 2D localization, the 1D correlation kernels in both directions were 11 elements long. The MBs were separated into six subsets based on three speed ranges and two flow directions (detailed in Section III.E).

The reconstruction was implemented in CUDA and run on a GPU RTX A6000 Ada (NVIDIA Corp., Santa Clara, CA, USA).

### D. Quantitative evaluation

To quantitatively compare the proposed fast projection-based and conventional full 3D processing, multiple evaluation metrics were employed. Structural similarity index (SSIM) [43] was utilized to evaluate the similarity between results obtained with different methods in both the clutter filtering (Figs. 2a-b) and the final ULM density maps (Figs. 5a and d). Linear regression analyses were performed to examine the relationships on estimated motions (Fig. 3d), and ULM speeds (Fig. 5j), with Pearson correlation coefficient $r$ and coefficient of determination $R^2$ used to quantify the correlation strength. Localization performance was quantitatively evaluated by comparing MB detection accuracy and missing rate [44] (Fig. 4b). Spatial resolution was estimated using the Fourier Shell Correlation (FSC) [45], with the resolution defined as the intersection of the FSC curve and the half-bit threshold (Fig. 5g). Computational efficiency was evaluated by averaging the runtime of each step over six repeated runs (Figs. 1a, 2d, 3c, 4f).

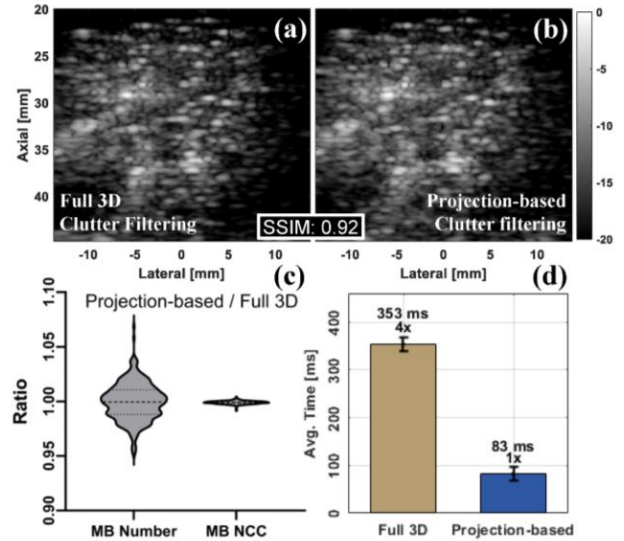

Fig. 2. Comparison of full 3D and projection-based clutter filtering. MIP of (a) full 3D and (b) projection-based clutter filtered volumes (central 6.8 mm in elevational direction). (c) ratio of MBs localized after projection-based to those localized after full 3D clutter filtering, and corresponding NCC ratio for 300 volumes. (d) Computational time comparison.

## III. RESULTS

The proposed projection-based framework consists of projection-based beamforming, clutter filtering, motion estimation, and localization. Since two-stage beamforming has been shown to introduce <1% error compared with 3D beamforming with 65× fewer computations [33, 34], its performance was not further evaluated here. In this Results section, the remaining steps are individually compared with conventional 3D processing, followed by qualitative and quantitative comparisons of the resulting ULM volumes.

### A. Clutter Filtering Results

Figures 2a and 2b show the MIP of MB volumes extracted using full 3D and projection-based clutter filtering, respectively. High similarity between the two is observed and supported by an SSIM of 0.92. Full 3D NCC-based localization was performed on an ensemble of 300 volumes, with the ratio of the number of MBs localized after projection-based clutter filtering to those localized after full 3D clutter filtering, and their corresponding NCC value ratio (reflecting localization quality [18]) summarized in Fig. 2c. The 25th and 75th percentiles of the localized MB ratio are 0.9882 and 1.0106, respectively, while the NCC ratio percentiles are 0.9980 and 0.9998, indicating very limited influence on the localization for most volumes. Fig. 2d shows a four-fold acceleration.

### B. Motion Estimation Results

Figure 3a shows the phantom results. RCA was translated at preset speeds of 0.4, 0.6, 0.8, 1.0, 1.2 and 1.4 mm/s using a motorized stage. Probe speed was estimated by linear fitting of the measured displacements, with the slope representing speed. Both full 3D and projection-based methods correlate perfectly with the preset speeds ($R^2 = 1.0$). Figure 3b shows projection-based estimated motion over time at a preset of 1.0 mm/s. The black curve indicates the error relative to expected motion, peaking at ~0.06 pixels near half-pixel displacements, which is a typical feature of sub-pixel estimation. Figure 3c compares the computational times, showing that the projection-based method accelerates 3D motion estimation by ~8-fold compared with full 3D method (38 ms vs. 286 ms; per 300 volumes).

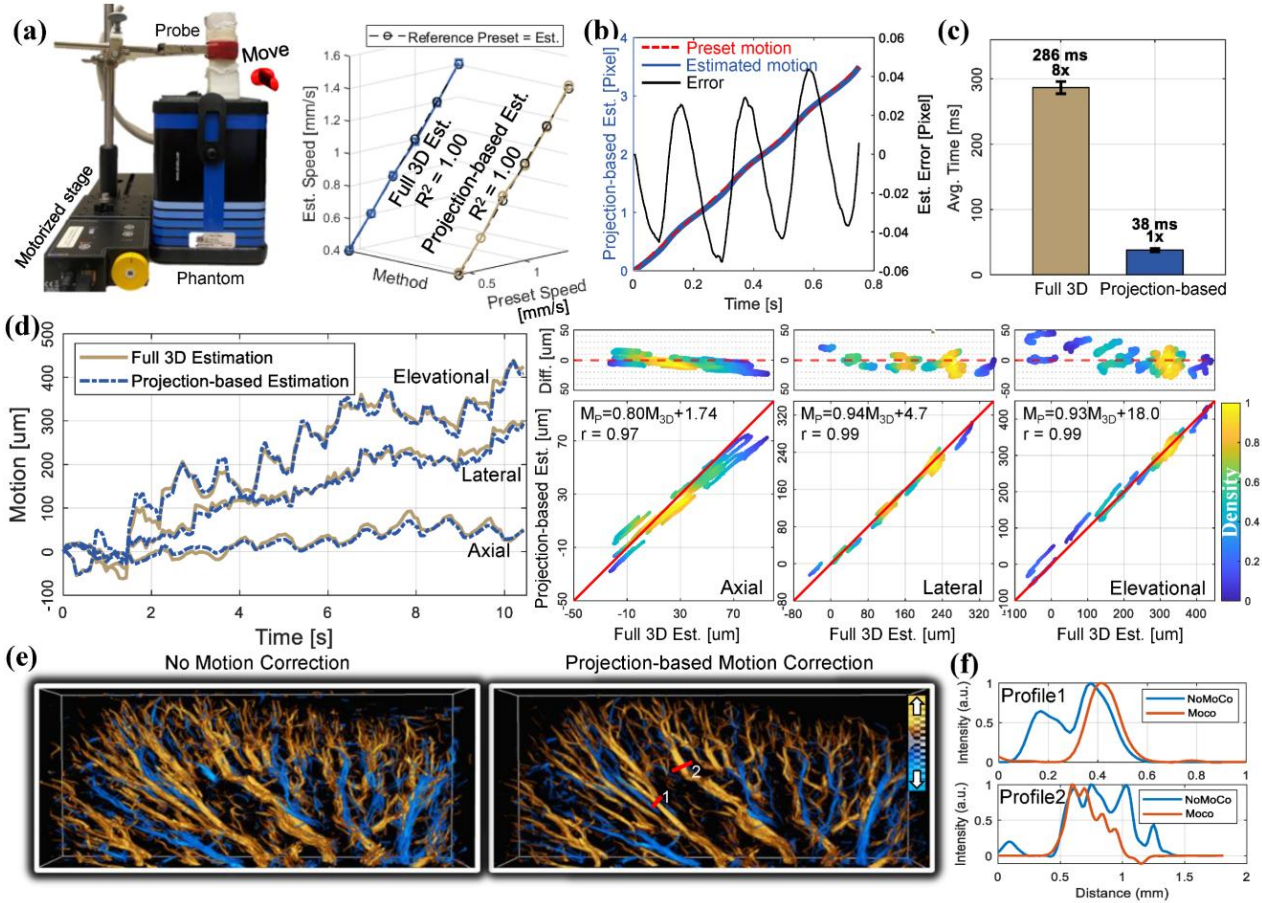

Fig. 3. Comparison of full 3D and projection-based fast motion estimation and quantification. (a) Phantom experiment setup (left). The RCA was clamped and translated at 0.4, 0.6, 0.8, 1.0, 1.2 and 1.4 mm/s using a motorized stage. Both full 3D and projection-based speed estimates correlate well with preset speed, with $R^2$ of 1.0 (right). (b) Projection-based motion estimates and preset motion over time, and corresponding estimation error. (c) Computational time comparison between full 3D and projection-based methods. (d) In vivo pig kidney motion estimated with full 3D and projection-based methods (left); density scatter plot with linear regression and error distribution (right). (e) 3D renderings of the ROI without motion correction and with the proposed projection-based motion correction. (f) Profiles corresponding to the red lines in (e).

Figure 3d presents the *in vivo* motions of pig kidney along axial, lateral and elevational directions estimated using the two methods. Note that B-mode-based motion estimation was performed within each ensemble, and those ensembles were further registered with their corresponding ULM volume (left). The scatter plots (right) demonstrate strong correlation between the full 3D and proposed projection-based estimations, with $r = 0.97$–0.99. Difference plots (above) confirm that estimation errors are mostly below 50 μm, corresponding to one voxel in ULM reconstruction.

Figure 3e presents 3D renderings of the cortex reconstructed without motion correction and with the proposed projection-based motion correction. The correction successfully aligns previously misregistered MB tracks, improving vessel delineation. Figure 3f shows profiles along the red lines, further illustrating the effectiveness of the proposed correction.

### C. MB Localization Results

Figure 4a shows representative MIP images of MBs randomly distributed and simulated with Field II at densities of 0.01 (top) and 0.1 (bottom) MB/mm³. Both full 3D and the proposed projection-based methods accurately localize MBs at low density but fail for closely spaced MBs (white arrows). Quantitative evaluation of localization performance across densities ranging from 0.002 to 0.1 MB/mm³ (Fig. 4b; error bars represent the standard deviation) indicates that projection-based localization achieves higher MB detection accuracy (orange arrow in Fig. 4a) but also a higher missing rate (green arrow), with both methods degrading at similar rates as density increases. Figure 4c shows representative MIPs of reconstructed speed maps in tube simulation at MB density of 0.06 MB/mm$^3$. While full 3D localization better populates voxels within the tube (lower MB missing rate, white arrows), it slightly overestimates (lower MB detection accuracy) flow speed (mean: 20.22 mm/s) compared to projection-based localization (mean: 20.08 mm/s) and the preset speed (20 mm/s). Mean flow speed quantification of all tracks under varying MB densities (Fig. 4d) shows a gradual drift above the preset speed and increasing standard deviation, demonstrating the impact of MB density on speed measurement accuracy. Similar localization performance was further confirmed *in vivo* using pig kidney data, with the mean ratio of localized and tracked MBs of projection-based method to full 3D method per ensemble ranging from 0.95 to 1 (Fig. 4e). Figure 4f shows the computational time for a single ensemble subset, comparing the proposed projection-based localization using separable correlation (2D Conv1d) with 3D localization using either separable (3D Conv1d) or full correlation (3D Conv3D). The comparison includes MB separation, enhancement, localization, and tracking as stacked bars. Using separable correlation in 3D localization achieves a 4.4× speedup over full 3D correlation, and the proposed 2D localization provides an additional 4.6× acceleration.

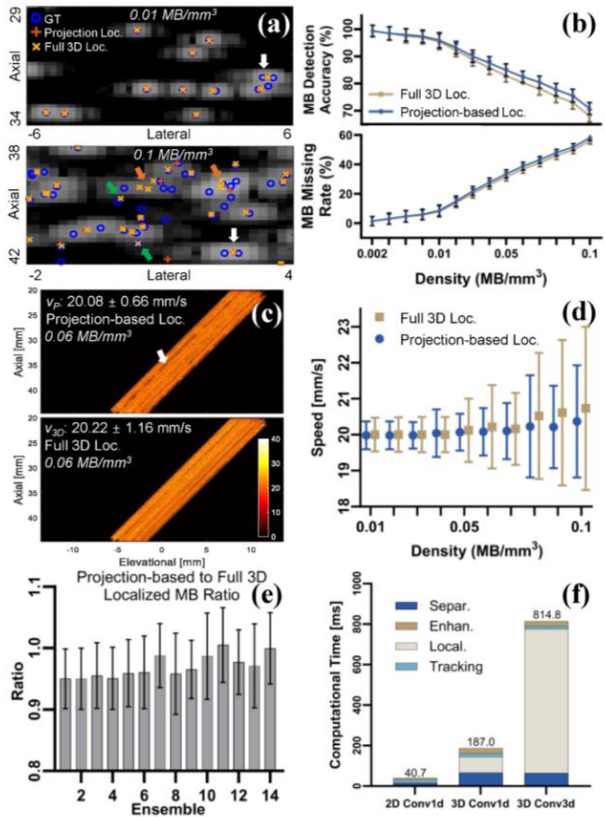

Fig. 4. Comparison of full 3D and projection-based 3D localization and quantification. (a) Representative projection images of MBs at densities of 0.01 and 0.1 MB/mm$^3$. (b) Detection accuracy and missing rate across varying MB densities. (c) Representative reconstructed flow speed projection at 0.06 MB/mm$^3$. (d) Statistical results of mean speeds of valid tracks as a function of MB density. (e) ratio of the number of MBs localized using projection-based to those localized using full 3D methods per ensemble in the *in vivo* pig kidney data. (f) Comparison of computational times for one MB subset of an ensemble.

### D. ULM Results
#### 1) Density Map

Figure 5a presents 3D renderings of contrast-enhanced XDoppler [40], full 3D processed ULM (hereafter referred to as *Full-3DULM*) and fast projection-based ULM (hereafter referred to as *Fast-3DULM*). Compared with contrast-enhanced XDoppler, both approaches provide higher resolution and clearer visualization of small cortical vessels. The two reconstructions appear visually similar, with a high SSIM of 0.93 confirming their agreement, as further illustrated in the zoomed-in view (Fig. 5b). Nevertheless, *Full-3DULM* shows slightly better reconstruction of large vessels with higher MB densities (white arrows).

Figure 5c zooms into the ROI indicated in Fig. 5b, comparing sub-volumes reconstructed with only 1.5 s (top) and full 10.5 s data (bottom) using *Full-3DULM* (left) and *Fast-3DULM* (right) processing. Quantifications over this ROI are shown in Fig. 5d. The vessel densities (ratio of vessel voxels to the total voxels in the volume) as a function of accumulated time are calculated and subsequently fitted to an exponential saturation model [18]. The vessel filling percentages (calculated as the ratio of vessel densities to the fitted maximum vessel density in the volume) are demonstrated in Fig. 5d, (top), reaching 89.4% and 91.8% after accumulating all the 14 ensembles with fitted maximum vessel densities of 15.2% and 16.3% for *Full-3DULM* and *Fast-3DULM* processing, respectively. The fitted filling curves of the two methods closely match across accumulation. The SSIMs between full 3D- and projection-based reconstructed sub-volumes (Fig. 5d, bottom) remain consistently high (> 0.88) during accumulation, with the slight decrease likely caused by localization difference-induced voxel misalignments.

Signed projections of the full 10.5 s reconstructions along the white dashed arrow in Fig. 5c are shown in Fig. 5e. The line profiles in Fig. 5f demonstrate that both *Full-* and *Fast-3DULM* resolve small vessels undetectable in XDoppler, with high agreement between their profiles. FSC analysis (Fig. 5g) yields a spatial resolution of 111 μm for both methods, below half the wavelength (148 μm), though potentially limited by the ULM voxel size of 50 μm used in this study.

#### 2) Velocity Map

Figure 5h presents the 3D rendering of the velocity maps obtained from *Full-* and *Fast-3DULM*, with zoomed-in views shown in Fig. 5i (top). Although the velocity distributions are visually similar, both global and zoomed views reveal a slight underestimation of flow speed in large vessels reconstructed using *Fast-3DULM*, also reflected in the histograms (bottom). For *Full-3DULM*, positive and negative velocities are centered at 34.8 mm/s and -27.8 mm/s, with mean values of 38.94 mm/s and -34.46 mm/s, respectively; for *Fast-3DULM*, the corresponding centers are 32.9 mm/s and -27.8 mm/s, with mean velocities of 38.87 mm/s and -34.98 mm/s. Overall, the velocity distributions obtained from these methods agree well, confirming their consistency. Voxel-wise comparisons (Fig. 5j;

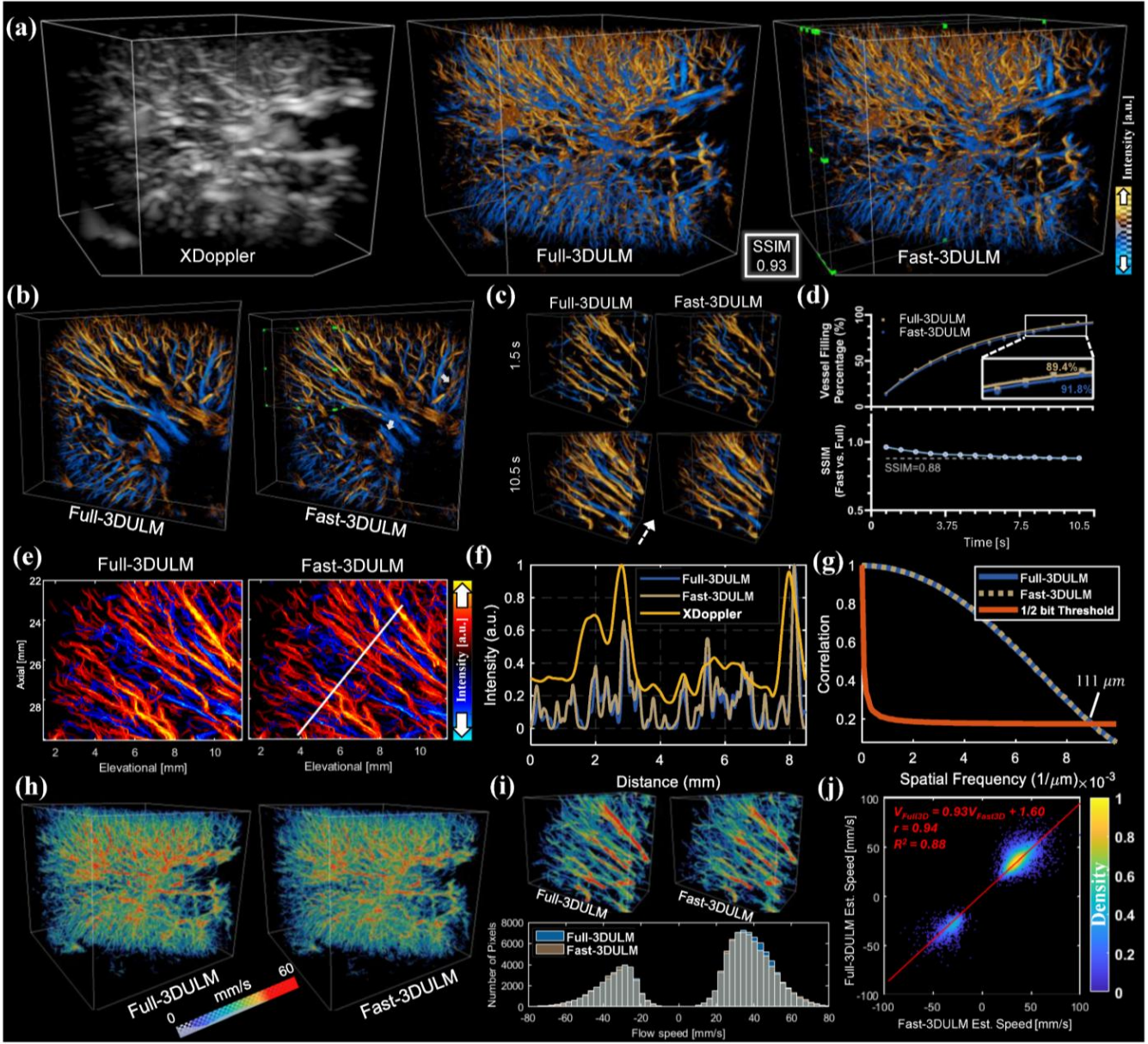

Fig. 5. ULM density and velocity maps obtained using full 3D (*Full-3DULM*) and proposed projection-based (*Fast-3DULM*) processing, along with quantification results. (a) 3D rendering of the full field-of-view microvascular network reconstructed using contrasted enhanced XDoppler, *Full-3DULM*, and *Fast-3DULM*. (b) Zoomed-in view of the ROI indicated in (a). (c) Zoomed-in ULM images reconstructed using 1.5 s (top) and 10.5 s (bottom). (d) Quantification within the ROI in (c): vessel filling percentage as a function of accumulation time (top) and SSIM between the *Full*- and *Fast*3DULM (bottom). (e) Signed MIP along the white dotted arrow of the 10.5 s accumulated ULM sub-volumes from (c). (f) Intensity profiles along the white line in (e) for XDoppler, *Full*- and *Fast*-3DULM. (g) FSC computed from the *Full*- and *Fast*-3DULM volumes. (h) Velocity maps reconstructed with *Full*- and *Fast*-3DULM. (i) Zoomed-in velocity maps corresponding to (c) (top), and their velocity distributions (bottom). (j) Voxel-wise comparison between velocities estimated with *Full*- and *Fast*-3DULM.

restricted to voxels with the same flow direction to avoid misalignment artifacts) show strong agreement between estimates, with linear slope of 0.93, $r = 0.94$, and $R^2 = 0.88$.

### E. ULM Results with Varying MB separation strategies

Figure 6a shows signed maximum projections of the same ROI in Fig. 5e reconstructed using different MB separation strategies. Increasing the number of MB subsets progressively enhances the reconstructed vasculature: the downward-flowing vessel (white arrow) is visible only after separating MBs by direction (Up1/Down1) and becomes clearer with further separation into two speed ranges (Up2/Down2). Separation into three speed ranges (Up3/Down3) reveals additional vessels (green arrows) visible only under this condition. These trends are confirmed by 3D vessel density quantification prior to projection (Fig. 6b), which increases with the number of subsets. Notably, the proposed *Fast-3DULM* and *Full-3DULM* yield very similar vessel densities across all separation strategies, confirming comparable performance. The continuously increasing vessel density highlights the importance of MB separation in 3D ULM, therefore, the results in Fig. 5 were obtained using Up3/Down3 strategy. The computational time of these methods differs in the MB separation, localization, and tracking step, as shown in Fig. 4f, which scales proportionally with the number of MB subsets. Further separation was not applied, as it may split continuous tracks across different subsets, complicating tracking and increasing computational cost.

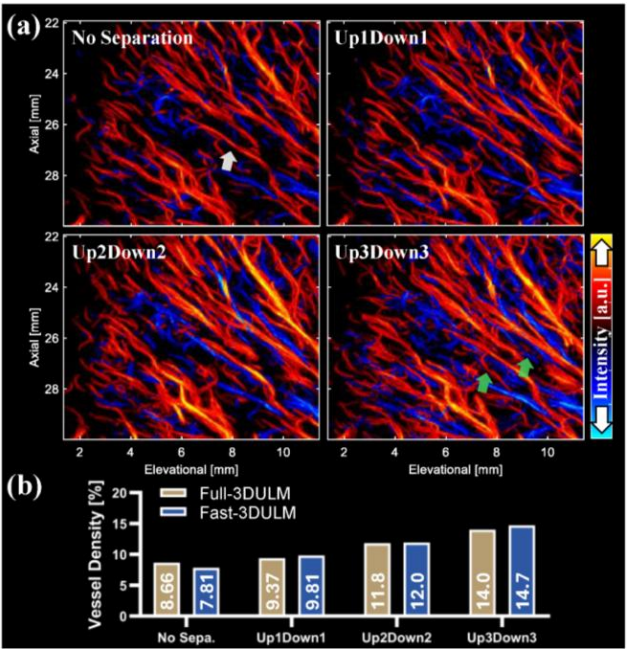

Fig. 6. (a) Signed MIP of the same 3D ROI as in Fig. 5e processed in 2D, using no MB separation and three separation strategies: two subsets by direction (Up1Down1), four subsets by direction and two speed ranges (Up2Down2), and six subsets by direction and three speed ranges (Up3Down3). (b) Vessel density within the 3D ROI for each method.

## IV. Discussion

3D ULM offers unique advantages for visualizing comprehensive microvasculature *in vivo*. However, its clinical implementation remains severely hindered by the large data volume required for adequate microvessel formation and the associated extremely heavy computational demands and extensive post-processing time. Therefore, developing fast and efficient 3D ULM methods is essential in clinical setup. Fast processing offers immediate visualization of microvasculature and hemodynamics to support timely and reliable diagnostic and interventional decisions. Here, we propose a projection-based processing framework for 3D ULM that achieves reconstruction faster than the data acquisition rate. Specifically, with each ensemble acquired in 0.75 s at 400 Hz, reconstruction can be completed within 0.52 s (Fig. 1a), allowing extra time for advanced processing or 3D rendering. Additionally, the proposed method maintains image quality comparable to full 3D method, demonstrating its potential for clinical translation.

Specifically, each of the key signal processing procedures in 3D ULM was decomposed into a series of 2D operations. For these steps, we demonstrated that the proposed frameworks achieved performance comparable to the conventional full 3D processing, as shown in Figs. 2-4. Combining all projection-based processing, the resulting ULM volumes of pig kidney achieved SSIM of 0.93 (Fig. 5a) and strong speed estimation agreement (Fig. 5j) with full 3D processing.

In this study, global SVD-based clutter filtering, rigid motion estimation and NCC-based MB localization were implemented as an illustrative example. As analyzed in the Method section, these steps scale linearly with the number of voxels. While effective for reconstructing kidney microvasculature, more advanced techniques, such as local SVD, non-rigid motion estimation, and optimization- or deep learning-based localization, could further improve robustness and image quality. Meanwhile, these approaches incur higher-than-linear computational cost, where even greater speed gains may be achieved by applying our projection-based strategy.

In our projection-based processing framework, the full 3D volume is divided into smaller blocks for localization. Using larger blocks increase the number of voxels along the projection direction, thereby reducing the number of projected 2D images to be processed and thus lower the overall computational cost; however, it also increases the likelihood of MB overlap, which may degrade localization accuracy. Therefore, an appropriate trade-off between reconstruction quality and computational efficiency should be considered when determining the block size. With the adoption of advanced algorithms capable of accurately localizing overlapping MBs, larger block sizes may become feasible, further improving computational efficiency.

In localization step, the 3D positions of MBs were recovered by pairing the 2D localizations obtained from two orthogonal projected slices, based on mutual minimal intensity ratios as an example. The matching accuracy can be further improved either by incorporating projections from additional directions, or by adopting better pairing strategies, such as weighing both the axial position difference and the intensity ratio.

This study aims to facilitate the clinical translation of 3D ULM. The pig kidney was chosen for method development and validation due to its close similarity to the human kidney in locating depth, size, vascular architecture, and perfusion characteristics. In clinical settings, a breath-hold of about 10 seconds is typically the maximum feasible duration, which is orders of magnitude shorter than acquisition times commonly used in ULM. To accommodate this constraint, we limited the reconstructed ULM voxel size to 50 μm (1/6 λ). A smaller voxel size would require longer acquisitions to adequately fill vessel voxels. Although larger voxel size reduces the achievable spatial resolution (111 μm in this study, exceeding the diffraction limit), it represents a practical compromise for wider clinical applications. Moreover, in clinical settings, breathing and probe drift make it challenging to achieve motion estimation accuracy smaller than the nominal voxel size of 50 μm selected in this study. To accumulate sufficient MB trajectories within 10 s, data were acquired at a relatively high MB density. This necessitates MB separation, which increases the computational cost approximately in proportion to the number of subsets, further highlighting the importance of efficient processing.

Currently, the raw data are collected and saved to disk during acquisition and processed offline. The data loading time was not included in the reported 0.52 s reconstruction time per ensemble. In the future, the proposed projection-based processing framework could be compiled as a *mexCUDA* file to run directly on the Verasonics scanner, enabling real-time performance evaluation and demonstrating its utility for providing immediate feedback during scanning to achieve more robust ULM imaging.